\documentclass[prb,aps,showpacs,showkeys,preprint]{revtex4}
\usepackage{graphicx}
\usepackage{amsmath}
\usepackage{amssymb}
\usepackage{amsthm}
\usepackage{booktabs}
\usepackage{stmaryrd}
\usepackage{longtable}
\usepackage[figuresright]{rotating}
\usepackage{multirow}
\usepackage{makecell}
\usepackage{dcolumn}
\usepackage{anysize}
\usepackage{color}
\marginsize{1.5cm}{1.5cm}{1.5cm}{1.5cm}
\newcolumntype{d}[1]{D{.}{.}{#1}}
\begin{document}
\title{Crystal field and magnetism of Pr$^{3+}$ and Nd$^{3+}$ ions in orthorhombic perovskites}
\author{P. Nov\'ak, K. Kn\'i\v{z}ek, M. Mary\v{s}ko, Z. Jir\'ak, and J.Kune\v{s}}
\affiliation{Institute of Physics of ASCR, Cukrovarnick\'a 10, 162 00 Prague 6, Czech Republic}
\date{\today}
\begin{abstract}
Fifteen  parameters characterizing the crystal field of rare-earth ions in the RMO$_3$ perovskites
(R = Pr, Nd, M = Ga, Co) are calculated by expanding the local Hamiltonian expressed in the basis of Wannier functions into a series
of spherical tensor operators. The method contains a single adjustable parameter that characterizes the
hybridization of R($4f$) states with the states of oxygen ligands.
Subsequently the energy levels and magnetic moments of trivalent R ion
are determined by diagonalization of an effective Hamiltonian which, besides the crystal field, contains
the $4f$ electron-electron repulsion, spin-orbit coupling and interaction with magnetic field. In
the Ga compounds the energy levels of the ground multiplet agree within few meV with those determined experimentally
by other authors. For all four compounds in question the temperature dependence of  magnetic susceptibility
is measured on polycrystalline samples and compared with the results of calculation.
For NdGaO$_3$ theory is also compared with
the magnetic measurements on a single crystal presented  by Luis {\it et al.} Phys. Rev. B {\bf 58}, 798 (1998).
A good agreement between the experiment and theory is found.
\end{abstract}
\maketitle
\section{Introduction}

The rare-earth elements can be naturally incorporated in the crystal structures of many
important groups of materials such as cuprates, manganites, or cobaltites to name a few.
While their $s$, $p$, $d$ electrons are involved in the bonding the $4f$ orbitals are
in many cases only weakly coupled to the rest of the crystal. This makes them
a potential local probe able to detect internal exchange fields. However, for the
rare-earth ions to serve this purpose knowing their response to the external field
is a necessity. At temperatures well above the Kondo temperature
the physics of the $4f$ electrons can be captured by an effective atomic
Hamiltonian, in which the crystalline environment is described by a set of crystal-field
parameters (CFP).

To determine the CFP experimentally the optical $f$-$f$ transitions
may be used to map out the individual crystal-field split multiplet
levels.~\cite{duan,gruber} The approach suffers from several limitations.
For example, the polarization analysis is required to orient the crystal field with
respect to the crystal axes or non-dipole transitions are necessary for
R ion in a centrosymmetric position. Most importantly the optical methods
are not applicable in metals where the $f$-$f$ transitions are hidden
by the optical response of the free carriers. Therefore theoretical
determination of CFP is of great importance.

Recently, some of us have developed a method to compute
CFP based on a first-principles electronic structure and the Wannier
projection.~\cite{novak1,novak2} The method was tested on the
orthorhombic aluminates R:YAlO$_3$ and TbAlO$_3$ for which extensive
optical data exist. For Nd:YAlO$_3$, Er:YAlO$_3$ and TbAlO$_3$ we compared
the calculated energy levels of R$^{3+}$ ion with the experimental ones
to get satisfactory agreement to within a few meV.

In this article we study the rare-earth gallates NdGaO$_3$ and PrGaO$_3$, and cobaltites
NdCoO$_3$ and PrCoO$_3$, for which
optical $f$-$f$ spectra are not available. Besides the energy levels, which
we compare to the inelastic neutron scattering data for NdGaO$_3$~\cite{podle_nd}
and PrGaO$_3$,~\cite{podle_pr}
we focus on the magnetic response of R ions and calculate the anisotropic
$g$-factors and van Vleck susceptibilities. These are compared to the
monocrystal data for NdGaO$_3$~\cite{luis} and to new susceptibility data
on polycrystals for all four compounds. We find an excellent match
between the lowest multiplet levels for NdGaO$_3$ and suggest
reassignment of the lowest multiplet levels to the incomplete
experimental set for PrGaO$_3$.~\cite{podle_pr} The anisotropic
$g$-factor and van Vleck susceptibility calculated for NdGaO$_3$ compare
well to the monocrystal data~\cite{luis}, while the polycrystal averaged
susceptibilities reproduce the corresponding temperature dependencies
for all studied materials.

The paper is organized as follows: in section \ref{sec:methods} the effective Hamiltonian method and
the way the crystal field is calculated are briefly
summarized, as well as the way to determine the magnetic moments. In section \ref{sec:exp}
the experimental details are given. In section \ref{sec:results} the experimental
and theoretical results and their comparison are presented. In section \ref{sec:discussion} the results are discussed, followed by
the conclusions in section \ref{sec:conclusions}.

\section{methods}
\label{sec:methods}
\subsection{Effective Hamiltonian}
The effective Hamiltonian operating on the $4f$ states can be written as
\begin{equation}
\label{eq:h}
\hat{H}_{eff}=\hat{H}_A + \hat{H}_Z + \hat{H}_{CF},
\end{equation}
where $\hat{H}_A$ is the spherically symmetric atomic Hamiltonian, $\hat{H}_Z$ corresponds to the Zeeman
interaction and $\hat{H}_{CF}$ is the crystal field term. In the Wybourne notation \cite{wybourne} $\hat{H}_{CF}$
has the form
\begin{equation}
\hat{H}_{CF} = \sum_{k=0}^{k_{max}}\sum_{q=-k}^k B_{q}^{(k)} \hat{C}_{q}^{(k)},
\label{eq:hcf}
\end{equation}
where $ \hat{C}_{q}^{(k)}$ is a spherical tensor operator of rank $k$ acting on the
$4f$ electrons of the R ion, for which $k_{max}$ is equal to six. The coefficients $B_{q}^{(k)}$ are
the crystal field parameters. Hermiticity of $\hat{H}_{CF}$ requires that $(B_{-q}^{k})^* = (-1)^q B_{q}^{k}$.
The details of $\hat{H}_A$ are given e.g. in Ref. \onlinecite{hufner}.
In the gallates and cobaltites with the orthorhombic $Pbnm$ structure (see Appendix A), the R cations are located on sites of $C_s$
 point symmetry, which leads to nine independent crystal field parameters, three are real ($k$ = 2, 4, 6,
 $q$ = 0) and six complex ($k$  = 2, 4, 6;  $q = 2,4,6; q \leq k$).

\subsection{Calculation of crystal field parameters}
The method to calculate the crystal field parameters may be divided into four steps.
\begin{enumerate}
\item {Electronic structure of the material is calculated with the $4f$ electrons treated as the core electrons.
Here we use the density functional based WIEN2k package \cite{wien}. The calculation is non spin-polarized, which
leads to the spin independent CFP. }
\item{The eigenvalue problem is solved with the potential determined in the first step.
The $4f$ orbitals are treated as the valence states, but only $2s$ and $2p$ states of oxygen are allowed to mix with
them. Mixing of $4f$ with other than oxygen states is prevented by shifting these states to a high energy, using an 
orbital potential.\cite{novak2}
The position of $4f$ states relative to oxygen states, which is not correctly described by any local or semilocal
density functional method, is adjusted by shifting the oxygen states by energy
$\Delta$ - to this end the orbitally dependent potential is used. $\Delta$ is an adjustable parameter, but in calculations reported here
its value was fixed at -0.6 Ry (-8.2 eV), which gave a good agreement between experiment and theory for Nd:YAlO$_3$.\cite{novak2}}
In the same reference the physics behind the parameter $\Delta$ is discussed in detail.
\item{The Bloch states of the $4f$ band and the local Hamiltonian $\hat{H}_{4f}$ is extracted.
$\hat{H}_{4f}$ is a seven by seven matrix which reflects
the symmetry of R site. For this step the wien2wannier \cite{kunes} and the  wannier90 \cite{wannier} programs are used.}
\item{The local Hamiltonian is expanded in the spherical tensor operators
using standard linear algebra program. The expansion coefficients are the CFP.}
\end{enumerate}

\subsection{Electronic structure, magnetic moments and susceptibility of rare-earth ion}
\label{sec:E_m}
To find the eigenvalues of $\hat{H}_{eff}$  modified 'lanthanide' program \cite{edwardsson} was used.
 The atomic parameters of $\hat{H}_A$ parameters are only weakly
material dependent and we used the values determined by Carnall {\it et al.} \cite{carnall} for R$^{3+}$
ions in LaF$_3$.

The spectrum of free Pr$^{3+}$ ion (electron configuration $4f^2$) consists of three $|L,S,J\rangle$ multiplets,
the lowest one being $^3H_4$ ($J=4$). The $4f^3$ configuration of Nd$^{3+}$
ion leads to four multiplets with the $^4I_{9/2}$ ground multiplet.
The crystal field of the $C_s$ symmetry
splits the Pr$^{3+}$ multiplets into 91 orbital singlets, while there are 182 Kramers doublet
states originating from the multiplets of Nd$^{3+}$. The energies of Kramers
doublets depend on the magnetic field as
\begin{equation}
\label{eq:tensors}
 \varepsilon_\pm =\varepsilon(0) \pm \frac{1}{2}\,\mu_B |\vec{B}|\, g(\vec{n}) -\frac{1}{2}\vec{B}\, \hat{\chi}^{vV}\vec{B};
\;\; \vec{n}=\frac{\vec{B}}{ |\vec{B}|},
\end{equation}
where $\varepsilon(0)$ is the energy in zero field, $g$ is the effective $g$-factor and $\hat{\chi}^{vV}$ is the
van Vleck susceptibility. The $g$-factor depends on the direction of the external magnetic field
$\vec{B}/|\vec{B}|$= ($\vartheta_x, \vartheta_y,\vartheta_z$) as given by
\begin{equation}
 g = \sqrt{g_{x}^{2} \vartheta_{x}^{2} + g_{y}^{2} \vartheta_{y}^{2} + g_{z}^{2} \vartheta_{z}^{2} }.
\end{equation}
Keeping with tradition we call $g_x$, $g_y$ and $g_z$ principal components of the $\hat{g}$-tensor, despite the
fact that $\hat{g}$ is not a true tensor, as discussed e. g. by Abragam and Bleaney.\cite{ab} The situation is
briefly clarified in the Appendix \ref{sec:g}.
For the non Kramers Pr$^{3+}$ ion the linear term in $B$ is missing, leading to a quadratic dependence on the
magnetic field.

 To determine $\hat{g}$ and $\hat{\chi}^{vV}$ tensors the eigenenergies
$\varepsilon_i$ were calculated in an external magnetic field and expanded  to the second power of $B$. The magnetic moments of the
eigenstates (in Bohr magnetons) for the field in the direction $\alpha$ are given by the field derivative of the eigenenergies:
\begin{equation}
 m_i^{(\alpha)} = -\frac{\mathrm{d}\varepsilon_i^{(\alpha)}}{\mathrm{d}B}.
\end{equation}

The low local symmetry admits a nondiagonal $\hat{g}$ and $\hat{\chi}^{vV}$ component $g_{ab}$ and $\chi_{ab}^{vV}$. In
order to determine the magnetic moment and susceptibility of a polycrystal the canonical form of the  $\hat{g}$-tensor
is needed. The principal $z$ axis of both tensors is parallel to the orthorhombic $c$ axis, while
for the two inequivalent R sites the axis $x$ makes an angle $\pm\alpha_g$ and  $\pm\alpha_{vV}$ with the orthorhombic $a$ axis.
To determine the canonical form of the tensors the  $\varepsilon_i(B)$ dependence was calculated
with the external field parallel to the orthorhombic axes and to the direction $\omega$ which makes angle
$\pi$/4 with the orthorhombic $a$ axis. From these four quantities
the canonical form of the tensor $\hat{g}$ may be obtained from
\begin{alignat}{2}
\label{eq:gxy}
 \alpha_g= & \frac{1}{2} \mathrm{arctg} \frac{2 g_{\omega}^2 - g_{aa}^2 - g_{bb}^2}{g_{aa}^2 - g_{bb}^2},\\
\nonumber g_x = & \sqrt{\frac{g_{aa}^2 + g_{bb}^2}{2} + \frac{g_{aa}^2 - g_{bb}^2}{2\cos(2\alpha_g)}},\;\;
g_y = \sqrt{\frac{g_{aa}^2 + g_{bb}^2}{2} - \frac{g_{aa}^2 - g_{bb}^2}{2\cos(2\alpha_g)}},\;\;g_z=g_{cc},
\end{alignat}
where the index of eigenstate was omitted. Analogous relations hold also for the susceptibility tensor - it is
only necessary to replace the squares of the $\hat{g}$-tensor components by the components of $\hat{\chi}^{vV}$.

To obtain the $g$-factor and the susceptibility of a polycrystal averaging over the polar and azimuthal angles was
performed. The temperature dependence of the R magnetic moment was calculated using the Boltzmann statistics:
\begin{equation}
M^{(\alpha)}(T) = \sum_i m_i^{(\alpha)} \exp(-\varepsilon_i^{(\alpha)}/kT)/\sum_i \exp(-\varepsilon_i^{(\alpha)}/kT),
\end{equation}
where $k$ is the Boltzmann constant and $\alpha$ denotes the direction of the external magnetic field.

\section{experimental}
\label{sec:exp}
The magnetic measurements were performed on polycrystalline samples prepared by a standard ceramic method.
 The study was carried out using a SQUID magnetometer MPMS-XL (Quantum Design) and included DC susceptibility
measured over the temperature range of $2-300$~K in applied field of 1 T.

\section{results}
\label{sec:results}
\subsection{Energy levels}
\begin{figure}
\includegraphics[width=12cm]{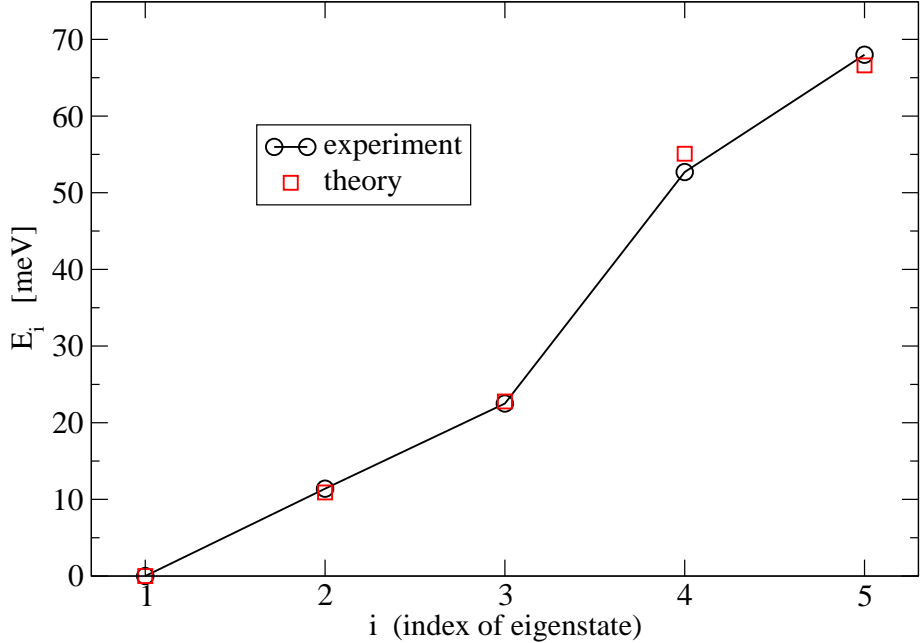}
\caption{(Color online) Splitting by crystal field of the $^4I_{9/2}$ multiplet of Nd$^{3+}$ ion in NdGaO$_3$.
Experimental data were taken from Ref. \onlinecite{podle_nd}. The broken line serves as a guide for eyes only.}
\label{fig:ndga_E}
\end{figure}

Values of the nine nonzero CFP calculated as described above are shown in Appendix \ref{sec:parameters}.
The $^4I_{9/2}$ ground multiplet   of the free Nd$^{3+}$ ion is split in the $C_s$ crystal field into five Kramers doublets.
Their energies for NdGaO$_3$  calculated  with the
CFP from Table \ref{tab:cfp} and atomic parameters of Carnall {\it et al.} \cite{carnall} are shown in Fig. \ref{fig:ndga_E}.
In Ref. \onlinecite{podle_nd} the energies of these doublets in NdGaO$_3$ were determined using the inelastic neutron scattering
and an attempt was made to interpret the experiment using the crystal field Hamiltonian. Unfortunately the CFP calculation was based on
questionable assumptions, it contained several fitting parameters and, as pointed out by Rudowicz and Qin \cite{rudowicz}, inconsistent
notation was used. Our results, based on the new set of CFP, agree very well with the experimental data as shown in Fig. \ref{fig:ndga_E}.

\begin{figure}
\includegraphics[width=12cm]{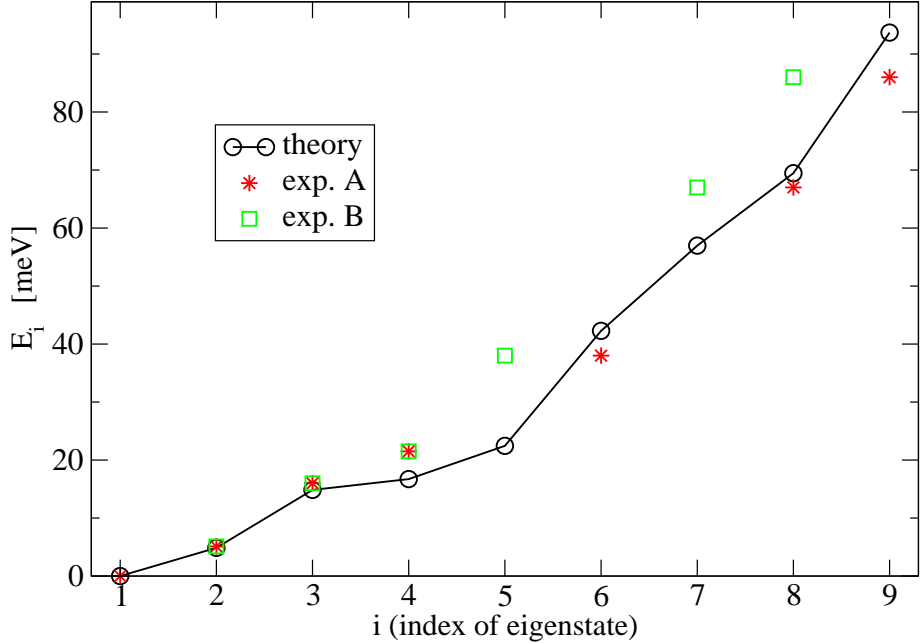}
\caption{(Color online) Splitting by crystal field of the $^3H_{4}$ multiplet of Pr$^{3+}$ ion in PrGaO$_3$.
 Experimental data exp. B (squares) were taken from Ref. \onlinecite{podle_pr}. See the text for explanation of the meaning of exp. A data. }
\label{fig:prga_E}
\end{figure}

\begin{figure}
\includegraphics[width=12cm]{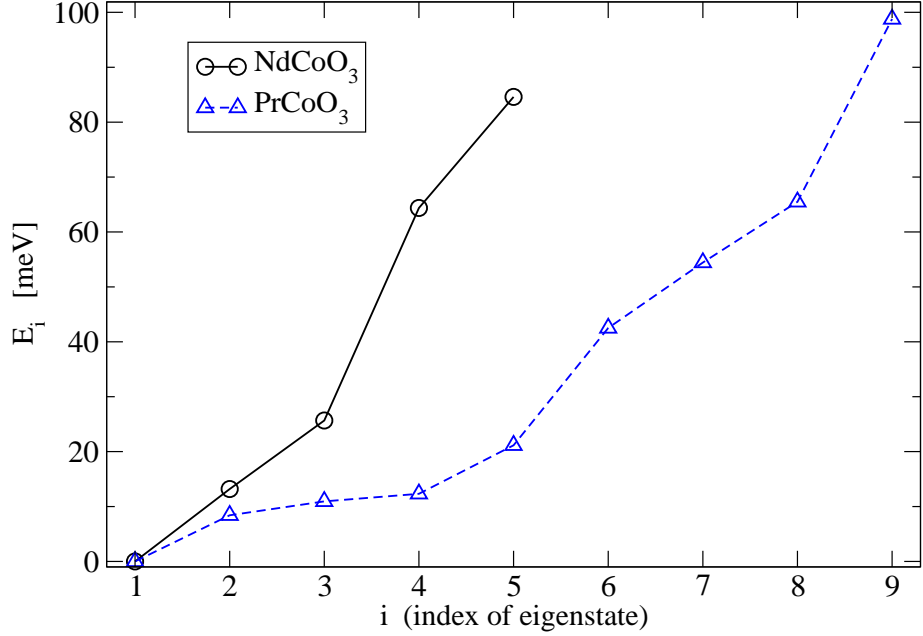}
\caption{(Color online) Calculated splitting by crystal field of the $^4I_{9/2}$ multiplet of  Nd$^{3+}$  and  the $^3H_{4}$ multiplet of Pr$^{3+}$ in RCoO$_3$.}
\label{fig:rco_E}
\end{figure}

The same authors applied  inelastic neutron scattering to study also PrGaO$_3$.\cite{podle_pr} The ground state $^3H_{4}$
multiplet of the Pr$^{3+}$ ion is  split by crystal field  into nine singlets, only six of the
excited singlets were detected experimentally, however. The analysis then poses a problem of identifying the undetected singlets.
Podlesnyak {\it et al.} decided that the sixth and the ninth singlet went undetected. The analysis suffers from the same problems as for
NdGaO$_3$ \cite{rudowicz}, however. In Fig. \ref{fig:prga_E} we compare our calculations with two possible assignments of
experimental data. Selecting the fifth and the seventh singlets
undetected (denoted exp. A) leads to a significantly better agreement with the experiment than the
choice of Podlesnyak {\it et al.} (exp. B).
There is no experimental information on energy levels of R$^{3+}$ ions in the cobaltites.
The present calculation indicates that the levels (Fig. \ref{fig:rco_E}) are not far from the levels in the gallates.
\subsection{Magnetic moments and susceptibility}
\label{sec:m_chi}
To fully characterize the magnetism of  Nd$^{3+}$ ion eight quantities for each eigenstate are needed:
three components of the $\hat{g}$ and $\hat{\chi}^{vV}$ tensors along the orthorhombic axes and a component along the $\omega$ direction
intersecting the angle between $a$ and $b$.
 In the case of the non Kramers Pr$^{3+}$ ion four components of $\chi^{vV}$ tensor are sufficient.
 For NdCoO$_3$, PrGaO$_3$ and PrCoO$_3$ these quantities
 are collected in the Appendix B. The data for NdGaO$_3$ are shown
in Table \ref{tab:ndga_e_g} and for the ground doublet they are compared to experimental data of Luis {\it et al}.\cite{luis}
 These authors measured the ac magnetic susceptibility of a NdGaO$_3$ single crystal along the three orthorhombic
axes in the temperature range 0.07 K $<  T  <$ 50 K in which only the lowest Kramers doublet is appreciably populated.
The results showed that the Nd moments order antiferromagnetically at $T_N \sim$ 1 K. By fitting the experimental
results in the temperature range 5-50 K  Luis {\it et al.}\cite{luis} obtained the data denoted as 'exp.' in Table  \ref{tab:ndga_e_g}.
\begin{table}
\caption{Nd$^{3+}$ ion in NdGaO$_3$. Energy of five Kramers doublets originating from the $^4I_{9/2}$, multiplet, $\hat{g}$ and $\hat{\chi}^{vV}$ tensor components along
the orthorhombic axes and $\omega$ direction. $\chi^{vV}$ is units of $\mu_B$/T. The experimental values were determined by
Luis {\it et al.} \cite{luis} and they refer to the ground doublet.}
\centering
\setcellgapes{3.5pt}
\makegapedcells
\begin{tabular}{cccccc|cccc}
\hline
\hline
doublet & $\varepsilon(0)$ [meV]&$g_{aa}$& $g_{bb}$& $g_{cc}$& $g_{\omega}$&$\chi^{vV}_{aa}$& $\chi^{vV}_{bb}$&$\chi^{vV}_{cc}$&$\chi^{vV}_{\omega}$ \\
\hline
 1 &  0.00 & 1.998 & 2.735 & 2.576 & 1.907 & 0.0215 & 0.0125 & 0.0116 & 0.0043 \\
exp. &   &  1.98(1) & 2.63(1)& 2.83(1)&   & 0.018(1)  &0.012(1)  & 0.008(1) &  \\
 2 & 10.90 & 2.671 & 2.296 & 0.974 & 2.048 &-0.0109 & 0.0012 & 0.0075 & 0.0129 \\
 3 & 22.77 & 2.883 & 1.397 & 2.965 & 2.981 & 0.0000 &-0.0006 &-0.0068 &-0.0068 \\
 4 & 55.09 & 3.271 & 4.212 & 1.165 & 5.262 &-0.0002 &-0.0038 &-0.0009 &-0.0021 \\
 5 & 66.61 & 2.347 & 1.662 & 3.816 & 1.437 &-0.0084 &-0.0072 &-0.0093 &-0.0064 \\
\hline
\hline
\end{tabular}
\label{tab:ndga_e_g}
\end{table}

\begin{figure}
\includegraphics[width=12cm]{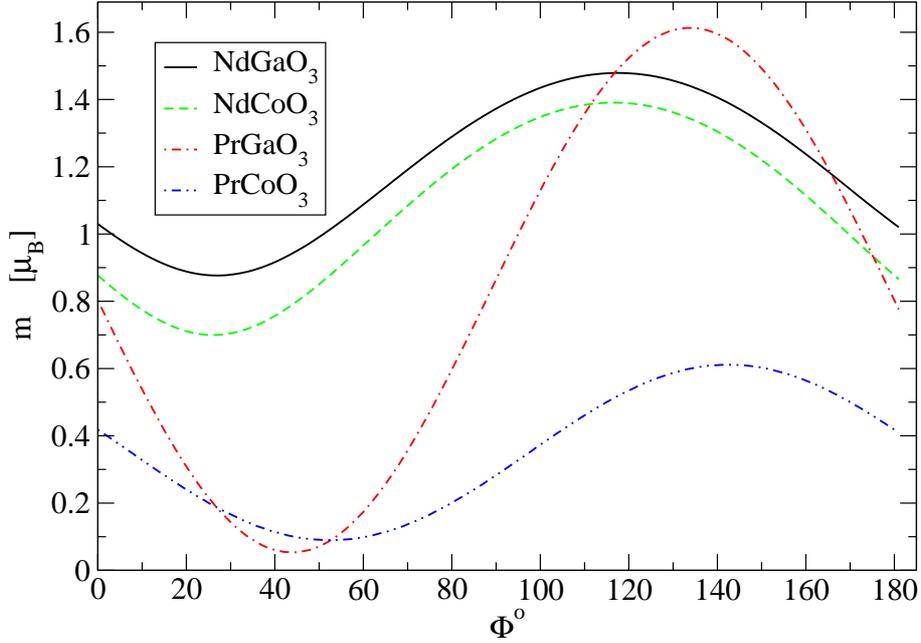}
\caption{(Color online). Angular dependence of Nd$^{3+}$ and Pr$^{3+}$ ground state magnetic moment.
External magnetic field 9 T is in the $c$ plane, $\Phi$ = 0$^\circ$ and $\Phi$ = 90$^\circ$ correspond to $a$ and $b$ axes, respectively. }
\label{fig:ab_ang}
\end{figure}

The anisotropy of the magnetic moment is demonstrated in Fig.  \ref{fig:ab_ang}, in which the angular dependence of the ground state
magnetic moment projected on the direction of the external magnetic field for all four compounds is shown.  The field
with the magnitude 9 T was confined to the $c$ plane.
Applying expressions (\ref{eq:gxy}) the values of  principal components of the $\hat{g}$-tensor and the angle $\alpha_g$ may be determined from
the components calculated in the orthorhombic coordinate system. In particular the data for NdGaO$_3$ (Table \ref{tab:ndga_e_g})
lead to the principal
components of $\hat{g}$-tensor $g_x$=1.734, $g_y$=2.910, $g_z$=2.576 and the angle of local
$x$-axis makes an angle $\alpha_g =\pm 25.1^{\,\mathrm{o}}$ with the orthorhombic $a$ axis (cf. also Fig. 9).
Similarly, for the $\hat{\chi}$-tensor the
values in Table \ref{tab:ndga_e_g} correspond to $\chi_\xi$=0.0354, $\chi_\eta$=0.0350, $\chi_\zeta$=0.0116 $\mu_B$/T
 and $\alpha_{vV} =\pm 56^{\,\mathrm{o}}$.

\begin{figure}
\includegraphics[width=12cm]{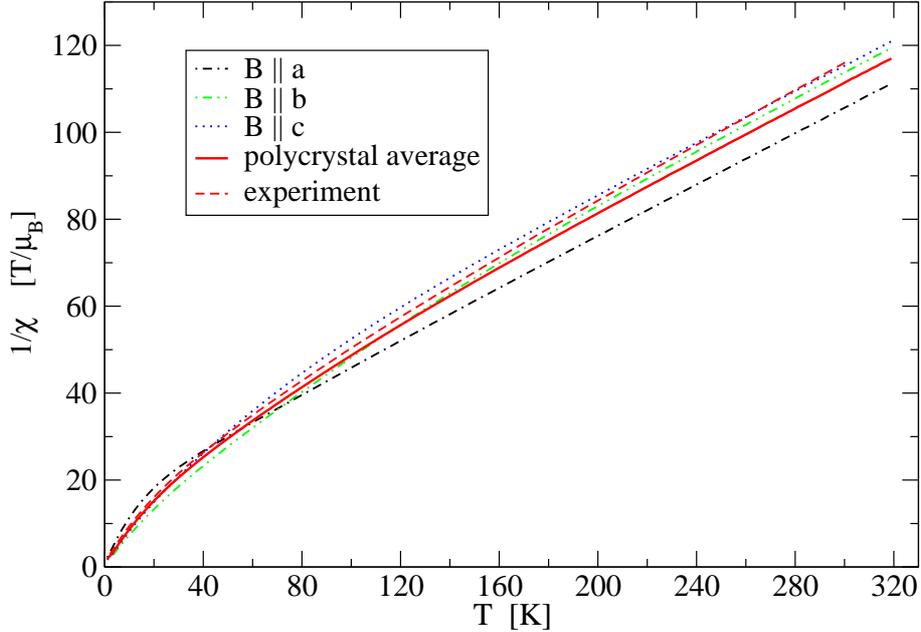}
\caption{(Color online) NdGaO$_3$. Temperature dependence of the inverse susceptibility. }
\label{fig:ndga_chi}
\end{figure}

\begin{figure}
\includegraphics[width=12cm]{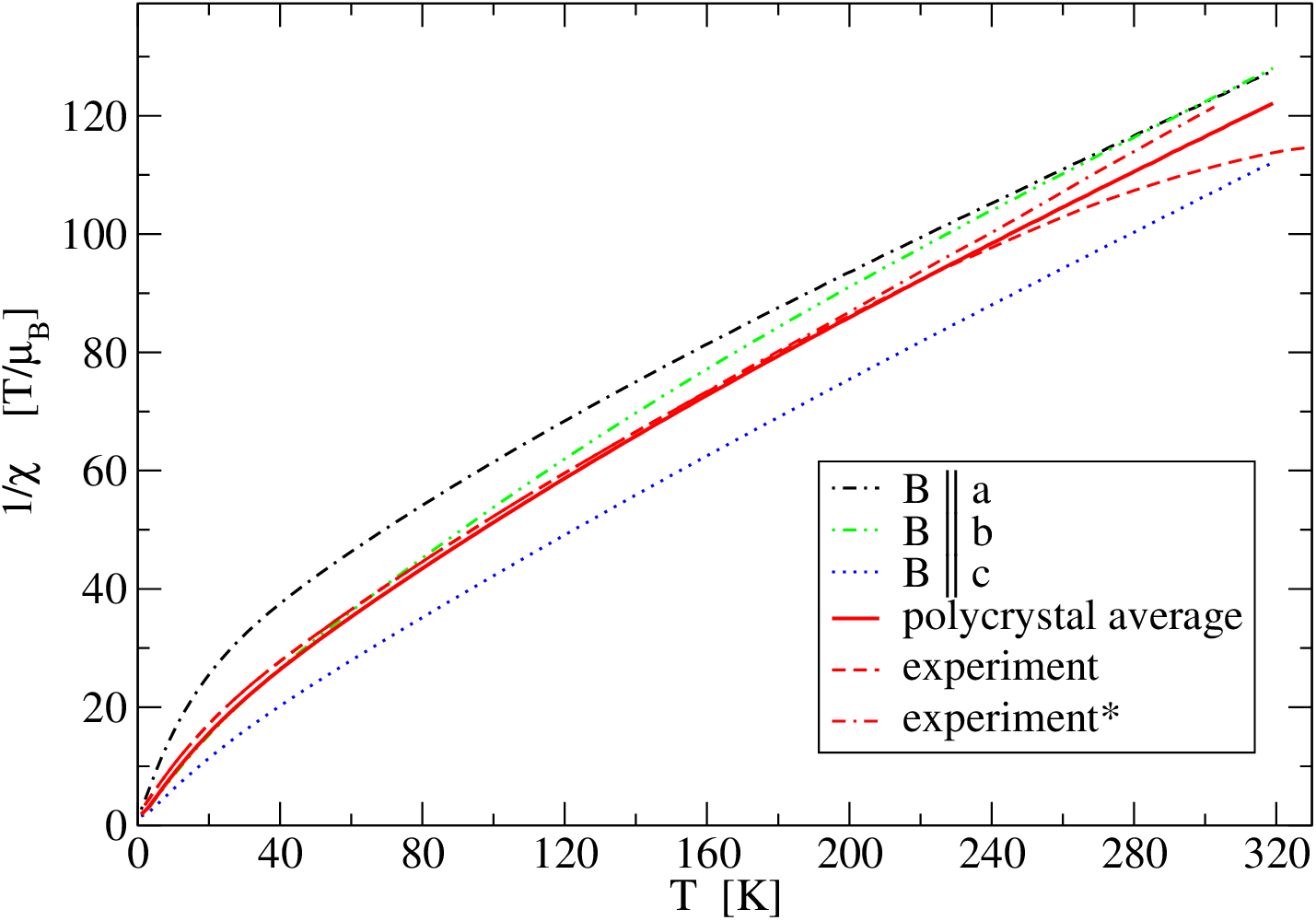}
\caption{(Color online) NdCoO$_3$. Temperature dependence of the inverse susceptibility.
Dot and dashed curve (experiment*) correspond to experimental data from which the contribution
of the high spin Co$^{3+}$ ions was subtracted (see section \ref{sec:discussion}). }
\label{fig:ndco_chi}
\end{figure}

\begin{figure}
\includegraphics[width=12cm]{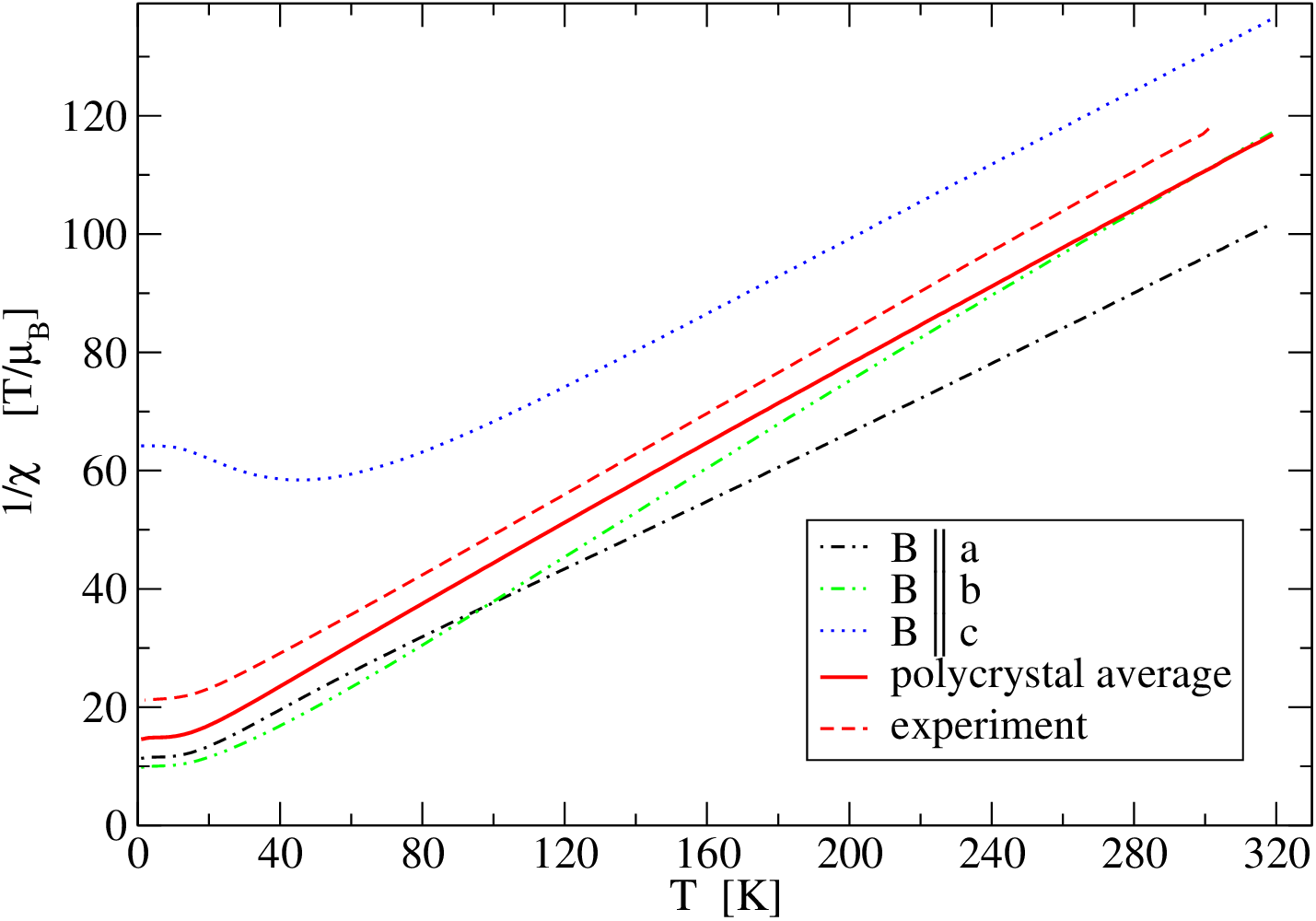}<
\caption{(Color online) PrGaO$_3$. Temperature dependence of the inverse susceptibility. }
\label{fig:prga_chi}
\end{figure}

\begin{figure}
\includegraphics[width=12cm]{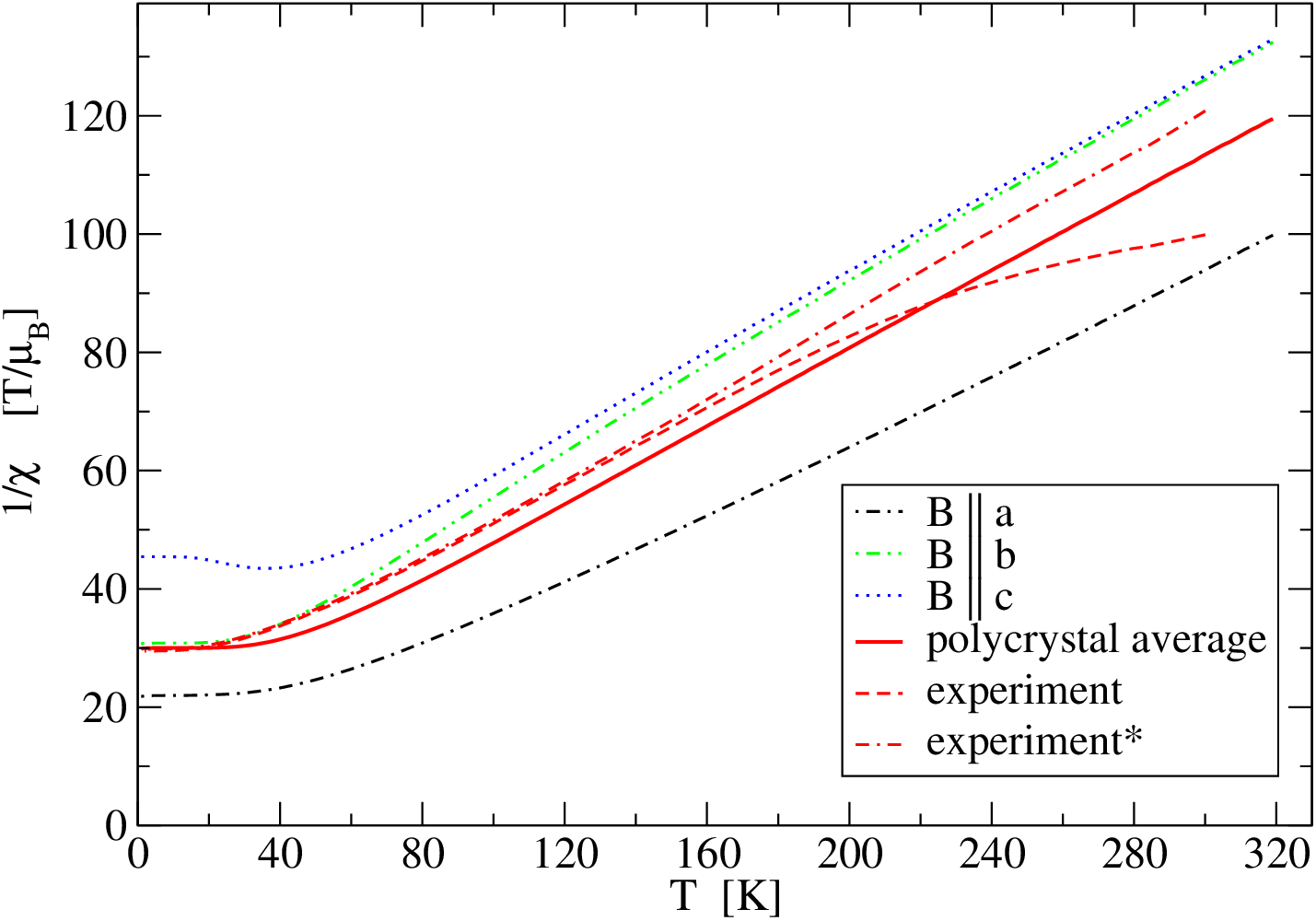}<
\caption{(Color online) PrCoO$_3$. Temperature dependence of the inverse susceptibility.
Dot and dashed curve (experiment*) correspond to experimental data from which the contribution
of the high spin Co$^{3+}$ ions was subtracted (see section \ref{sec:discussion}).}
\label{fig:prco_chi}
\end{figure}

In Figs. \ref{fig:ndga_chi} -  \ref{fig:prco_chi} the calculated temperature dependencies of the inverse susceptibilities in polycrystalline
samples are presented together with the experimental dependence for the four RMO$_3$ compounds.

\section{discussion}
\label{sec:discussion}
In RGaO$_3$ polycrystals both the experimental and calculated inverse magnetic susceptibilities are linear
in temperature starting from $~\sim$ 30 K and $\sim$ 120 K for R=Pr and Nd, respectively.
Deviation from the Curie-like behavior observed above $\sim$220 K for NdCoO$_3$ (Fig. \ref{fig:ndco_chi}) and
 above $\sim$150 K for PrCoO$_3$ (Fig. \ref{fig:prco_chi}) is
related to existence of the paramagnetic high spin states of Co$^{3+}$, lying close above the nonmagnetic low spin
ground state.  The corresponding contributions to the magnetic susceptibility
can be ascribed to the thermal excitation from the low-spin to the high-spin state.\cite{knizek,krapek}
To obtain the Curie-like behavior shown in Figs. \ref{fig:ndco_chi} and \ref{fig:prco_chi} the  excited states
of Co$^{3+}$ were assumed to lie at 130 meV and 95 meV for NdCoO$_3$ and PrCoO$_3$, respectively.

 There is a weak exchange interaction between the Nd$^{3+}$ moments, which
leads to the $C_z$-type antiferromagnetic ordering at the temperature $T_N \sim$ 1 K.\cite{luis} At low temperatures the magnetism of NdMO$_3$
is modified by the interaction also above $T_N$.  Using the mean-field approximation it would be straightforward
to include the exchange interaction in our scheme. The corresponding correction to the magnetic susceptibility decreases rapidly
with increasing temperature and it becomes insignificant above 40 K. The study of magnetism of the RMO$_3$ compounds
at low temperatures is in progress and will be the subject of a separate publication.
% as documented for the NdCoO$_3$ in Fig. \ref{fig:ndco_chi}.
%{\color{blue} ?? Zdenek doplnit molecular field approach + Figs M(B) zavislosti ??}

There are several approximation involved in the calculation of the CFP discussed in Ref. \onlinecite{novak2}.
Nevertheless, given that a single adjustable parameter is used, the agreement between the theory and experiment is very good.
In particular, the results obtained for energy levels in PrGaO$_3$
(Fig. \ref{fig:prga_E}) indicate strongly that the two levels undetected by the neutron experiment \cite{podle_pr} were misidentified
previously \cite{podle_pr}. Especially gratifying is the agreement of the van Vleck susceptibility $\chi^{vV}$ in
NdGaO$_3$ (Table I). Note that $\chi^{vV}$ was introduced by Luis {\it et al.} \cite{luis} to fit the susceptibility data
in the vicinity of the N\`{e}el temperature. The agreement thus provides theoretical justification of the
analysis of experimental data.

\section{conclusions}
\label{sec:conclusions}
We have calculated complete sets of crystal-field parameters for the rare-earth
ions in NdGaO$_3$, PrGaO$_3$, NdCoO$_3$ and PrCoO$_3$. These were used to determine
the anisotropic $g$-factors and van Vleck susceptibilities. The resulting
temperature dependencies of the susceptibility matches very well the experimental
data obtained on polycrystals. The calculated crystal-field split rare-earth levels
for NdGaO$_3$ and PrGaO$_3$ agree well with those obtained in
earlier inelastic neutron scattering experiments.~\cite{podle_nd,podle_pr}
Our results thus show that the R$^{3+}$ crystal field parameters in oxides can be calculated with
an accuracy which allows reliable prediction of the rare-earth magnetism.
This knowledge may serve as a basis for interpretation of experiments
in which the rare-earth ions serve as local probes of the magnetism
in cobaltites~\cite{jirak} as well as other materials.

\begin{acknowledgments}
The work was performed under the financial support of the Grant Agency of the Czech Republic within the Projects No.~204/11/0713,
Project No. P204/10/0284, and Project No.~13-25251S. We acknowledge Prof. H. Fujishiro of Morioka University, Japan for providing us with ceramic
samples of the gallium and cobalt perovskites.
\end{acknowledgments}

\appendix
\section{The orthoperovskite structure}
\begin{figure}
\includegraphics[width=12cm]{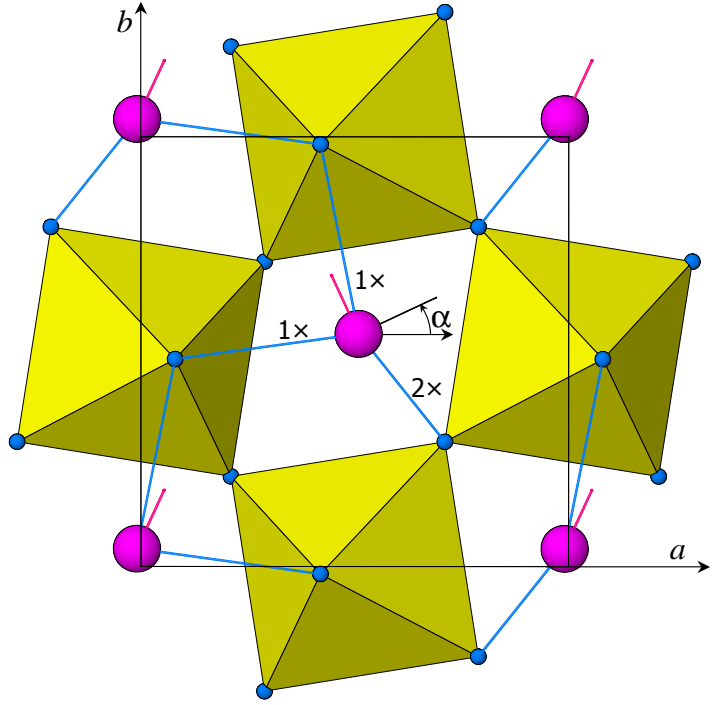}
\caption{Projection of NdGaO$_3$ structure along the $c$-axis. The rare-earth sites are located on
the mirror plane at $c/4$ together with oxygen atoms O1 at apical positions of the octahedra.
Other oxygen atoms O2 at basal positions are located close to the $c$=0 level. The shortest bonds
(two Nd-O1 and two Nd-O2 related by the mirror plane) are marked by the blue lines. The
direction in which lies the larger principal component of the $\hat{g}$-tensor in the $c$-plane,
derived from the data in Table~\ref{tab:ndga_e_g}, $g_y$ = 2.910 is depicted by red lines. The actual angle $\alpha_g$ is 25.1$^\mathrm{o}$
(see Fig. \ref{fig:ab_ang} and the text in section \ref{sec:m_chi}).}
 \label{fig:structure}
\end{figure}

The crystal structure of RMO$_3$ perovskites
(R = Pr, Nd, M = Ga, Co) consists of a pseudocubic array of the corner-shared MO$_6$ octahedra with the rare-earth cations in the cavities (see Fig. \ref{fig:structure}).
The orthoperovskite structure is characterized by a tilt of the octahedra by which the system responds to a size mismatch between the M and R cations. The regular
twelve-fold oxygen coordination of the R site known for the cubic perovskites is distorted in such a way that nine R-O distances decrease
and can be considered as bonding, while the four longest R-O distances gradually increase and are thus non-bonding. The resulting structure
possesses an orthorhombic $Pbnm$ symmetry which is described by four-times enlarged unit cell as shown by the crystallographic data summary in Table \ref{tab:structure}.

\begin{table}
\caption{The crystallographic data summary for RMO$_3$ compounds. Atom coordinates: Pr,Nd
$4c$(x,y,1/4), Co,Ga $4b$(1/2,0,0), O1 $4c$(x,y,1/4), O2 $8d$(x,y,z).}
\begin{tabular*}{\columnwidth}{@{\extracolsep{\fill}} lrrrr}
\hline
\hline
T [K]   & NdGaO$_3$ & NdCoO$_3$  & PrGaO$_3$ &PrCoO$_3$ \\
a [\AA] & 5.4276  &   5.3438    & 5.4557   &  5.3737    \\
b [\AA] &    5.4979 &  5.3345   &    5.4901 & 5.3395   \\
c [\AA] &  7.7078  &  7.5478    &    7.7275 &  7.5729    \\
\hline
x,Pr,Nd &   0.9909 &    0.9924  &   0.9926 &   0.9957   \\
y,Pr,Nd &    0.0414 &    0.0352   &   0.0352   &  0.0290    \\
x,O1    &   0.0800 &   0.0687   &  0.0758   & 0.0669    \\
y,O1    &    0.4826 & 0.4836  &  0.4848  & 0.4946    \\
x,O2    &    0.7107 &  0.7077    &  0.7132  &   0.7178  \\
y,O2    &  0.2903  &   0.2932   &     0.2871 &    0.2827   \\
z,O2    &    0.0422 &   0.0293   &    0.0404  &    0.0357  \\
\hline
\hline
\end{tabular*}
\label{tab:structure}
\end{table}

\section{Effective $g$-factor}
\label{sec:g}
We provide a brief derivation of Eq. (4). For detailed discussion of the
connection between the effective $g$-factor for Kramer's doublets with the
$g$-factor connecting the magnetic and angular momenta the reader is referred
to Ref.~\onlinecite{ab}. We start from the Zeeman Hamiltonian for the
Kramer's doublet
\begin{equation}
\hat{H}_Z=-\sum_{\alpha}\hat{m}^{\alpha}B_{\alpha},
\end{equation}
where $\hat{m}^{\alpha}$ are the Cartesian components of the
of magnetic moment operator. The eigenvalues of this $2\times2$ matrix
are
\begin{equation}
\varepsilon_{\pm}=\pm\sqrt{\sum_{\alpha\beta}\left(m^{\alpha}_{21}m^{\beta}_{12}-m^{\alpha}_{11}m^{\beta}_{22}\right)B_{\alpha}B_{\beta}}=
\pm \mu_B |\vec{B}|\, \sqrt{\sum_{\alpha\beta}G_{\alpha\beta}\vartheta_{\alpha}\vartheta_{\beta}},
\end{equation}
with $m^{\alpha}_{ij}$ being the matrix elements of operator $\hat{m}^{\alpha}$
between the states of the Kramer's doublet. We have used the fact that $m^{\alpha}_{11}+m^{\alpha}_{22}=0$ for each $\alpha=x,y,z$ due to the time reversal symmetry
between the states of the Kramer's doublet. Comparison with Eq.~\ref{eq:tensors}
gives the $g$-factor in the direction $(\vartheta_x,\vartheta_y,\vartheta_z)$ as
\begin{equation}
g=2\sqrt{\sum_{\alpha\beta}G_{\alpha\beta}\vartheta_{\alpha}\vartheta_{\beta}},
\end{equation}
where $\sum_{\alpha\beta}G_{\alpha\beta}\vartheta_{\alpha}\vartheta_{\beta}$ is a
real positive semi-definite quadratic form in the directional cosines.
The eigenvalues of $G_{\alpha\beta}$ are denoted $g_x^2$, $g_y^2$, and $g_z^2$.

\section{The values of parameters of the crystal field and magnetic tensors}
\label{sec:parameters}
The crystal field parameters obtained for the studied gallates and cobaltites are summarized in Table \ref{tab:cfp}.
They refer to the orthorhombic coordinate system of the $Pbnm$ structure.

\begin{table}
\caption{ Nonzero independent parameters of the crystal field in four compounds studied. All CFP are in units of meV.}
\centering
\setcellgapes{3.5pt}
\makegapedcells
\begin{tabular}{cccccc}
\hline
\hline
 k & q  &  NdGaO$_3$   &  NdCoO$_3$   &  PrGaO$_3$   &  PrCoO$_3$  \\
\hline
 2& 0& -29.19&         28.27&        -28.82&        -22.23       \\
 2& 2&  18.75+76.02i&   6.47+94.60i&  23.76+76.92i&  30.19+70.80i\\
 4& 0& -52.61&        -60.52       & -57.19       & -85.52       \\
 4& 2& -20.63+83.55i&  -9.92+120.83i& -18.52+85.55i& -15.88+95.57i\\
 4& 4&  20.76-75.35i&  46.45-64.44i&  23.71-63.86i&  44.12-48.34i\\
 6& 0& -74.44&       -121.36&        -80.84&       -104.66       \\
 6& 2&  11.97+34.56i&   8.78+63.53i&  10.11+36.99i&   9.02+47.49i\\
 6& 4&-164.72+0.52i&-195.90+6.00i&-175.20+0.33i&-216.61-1.57i\\
 6& 6&  15.46-1.95i&  11.63+0.48i&  15.48-1.51i&  16.52-2.33i\\
 \hline
\hline
\end{tabular}
\label{tab:cfp}
\end{table}

The energy, components of the $\hat{g}$ and $\hat{\chi}^{vV}$ tensors of five Kramers doublets of the Nd$^{3+}$ ion in NdGaO$_3$ were
summarized in Table \ref{tab:ndga_e_g}. Analogous data for NdCoO$_3$ are given in the Table \ref{tab:ndco_e_g}.
The energy and the components of the $\hat{\chi}^{vV}$ tensor of the nine orbital singlets of the Pr$^{3+}$ ion in  PrGaO$_3$ and PrCoO$_3$
are given in Table {\ref{tab:pr_e_g}.

\begin{table}
\caption{Nd$^{3+}$ ion in NdCoO$_3$. Energy of five Kramers doublets originating from $^4I_{9/2}$ multiplet, $\hat{g}$ and $\hat{\chi}^{vV}$ tensor components along
the orthorhombic axes and $\omega$ direction. Energy  $\varepsilon(0)$ is in meV, $\hat{\chi}^{vV}$ is in units of $\mu_B$/T. }
\centering
\setcellgapes{3.5pt}
\makegapedcells
\begin{tabular}{cccccc|cccc}
\hline
\hline
doublet & $\varepsilon(0)$ &$g_{aa}$& $g_{bb}$& $g_{cc}$& $g_{\omega}$&$\chi^{vV}_{aa}$& $\chi^{vV}_{bb}$&$\chi^{vV}_{cc}$&$\chi^{vV}_{\omega}$ \\
\hline
 1 & 0.00 &1.701 &2.560 &3.015 &1.442 &0.0149 &0.0106 &0.0118 &0.0024 \\
 2 &13.19 &1.773 &2.208 &2.432 &0.813 & -0.0072 & -0.0020 &0.0040 &0.0054 \\
 3 &25.66 &3.596 &2.524 &1.666 &3.800 &0.0026 &0.0018 & -0.0062 & -0.0013 \\
 4 &64.37 &2.964 &3.950 &1.659 &4.372 & -0.0012 & -0.0022 &0.0008 &0.0016 \\
 5 &84.60 &2.576 &2.152 &3.009 &1.605 & -0.0068 & -0.0057 & -0.0081 & -0.0063 \\
\hline
\hline
\end{tabular}
\label{tab:ndco_e_g}
\end{table}

\begin{table}
\caption{Pr$^{3+}$ ion in PrGaO$_3$ and PrCoO$_3$. Energy of nine singlets originating from $^3H_{4}$ multiplet, $\hat{\chi}^{vV}$ tensor components along
the orthorhombic axes and $\omega$ direction.  Energy  $\varepsilon(0)$ is in meV, $\chi^{vV}$ is in units of $\mu_B$/T. }
\centering
\setcellgapes{3.5pt}
\makegapedcells
\begin{tabular}{cccccc|ccccc}
\hline
\hline
& \multicolumn{5}{c}{Pr$^{3+}$ in PrGaO$_3$} & \multicolumn{5}{c}{Pr$^{3+}$ in PrCoO$_3$} \\
singlet&$\varepsilon(0)$ &$\chi^{vV}_{aa}$& $\chi^{vV}_{bb}$&$\chi^{vV}_{cc}$&$\chi^{vV}_{\omega}$&
$\varepsilon(0)$ &$\chi^{vV}_{aa}$& $\chi^{vV}_{bb}$&$\chi^{vV}_{cc}$&$\chi^{vV}_{\omega}$ \\
\hline
 1 & 0.00 & 0.0862 & 0.0991 & 0.0156 & 0.0062 & 0.00 & 0.0455 & 0.0324 & 0.0220 & 0.0107 \\
 2 & 4.45 &-0.0804 &-0.0814 & 0.0233 & 0.0065 & 8.41 &-0.0184 & 0.0343 & 0.0706 & 0.0083 \\
 3 &14.49 & 0.3206 & 0.0173 & 0.0316 & 0.2555 &10.94 & 0.4076 &-0.0473 & 0.0074 & 0.2444 \\
 4 &16.04 &-0.2543 & 0.0180 &-0.0177 &-0.2545 &12.32 &-0.4046 & 0.0243 &-0.0653 &-0.2488 \\
 5 &21.42 &-0.0591 &-0.0366 &-0.0355 & 0.0003 &21.15 &-0.0188 &-0.0233 &-0.0172 & 0.0009 \\
 6 &41.03 & 0.0111 & 0.0033 & 0.0095 & 0.0234 &42.51 & 0.0147 &-0.0008 & 0.0106 & 0.0232 \\
 7 &54.98 &-0.0185 &-0.0011 & 0.0050 &-0.0225 &54.43 &-0.0183 & 0.0045 & 0.0007 &-0.0179 \\
 8 &67.04 & 0.0082 & 0.0071 &-0.0138 & 0.0197 &65.49 & 0.0033 &-0.0069 &-0.0145 & 0.0019 \\
 9 &89.35 &-0.0118 &-0.0234 &-0.0157 &-0.0324 &98.78 &-0.0086 &-0.0147 &-0.0117 &-0.0201 \\
\hline
\hline
\end{tabular}
\label{tab:pr_e_g}
\end{table}

\end{document}